\documentclass[preprint,showpacs,preprintnumbers,amsmath,amssymb]{revtex4}

\usepackage{graphicx}
\usepackage{dcolumn}
\usepackage{bm}
\usepackage{amssymb}

\begin{document}
\newcommand{\vecx}{\mbox{\boldmath $x$}}
\newcommand{\vecp}{\mbox{\boldmath $p$}}
\newcommand{\vecr}{\mbox{\boldmath $r$}}

\title{Specific heat and entropy
of $N$-body nonextensive systems
}

\author{Hideo Hasegawa}
\altaffiliation{hideohasegawa@goo.jp}
\affiliation{Department of Physics, Tokyo Gakugei University,  
Koganei, Tokyo 184-8501, Japan}%

\date{\today}

\begin{abstract}
We have studied finite $N$-body $D$-dimensional nonextensive
ideal gases and harmonic oscillators, 
by using the maximum-entropy methods with the $q$- and normal averages
($q$: the entropic index).
The validity range, specific heat and Tsallis entropy obtained by the two
average methods are compared.
Validity ranges of the $q$- and normal averages are 
$0 < q < q_U$ and $q > q_L$, respectively,
where $q_U=1+(\eta DN)^{-1}$, $q_L=1-(\eta DN+1)^{-1}$ and
$\eta=1/2$ ($\eta=1$) for ideal gases (harmonic oscillators).
The energy and specific heat in the $q$- and normal averages coincide with
those in the Boltzmann-Gibbs statistics, 
although this coincidence does not hold for the fluctuation of energy. 
The Tsallis entropy for $N \vert q-1\vert \gg 1$ obtained by the $q$-average  
is quite different from that derived by the normal average,
despite a fairly good agreement of the two results for $\vert q-1 \vert \ll 1$.
It has been pointed out that first-principles approaches
previously proposed in the superstatistics yield $additive$ $N$-body entropy 
($S^{(N)}= N S^{(1)}$) which is in contrast with the $nonadditive$ Tsallis entropy.

\end{abstract}

                             
        


\maketitle
\newpage

\section{Introduction}
Many studies have been reported on the nonextensive thermodynamics
since Tsallis proposed the so-call Tsallis entropy defined by 
\cite{Tsallis88}-\cite{Martinez00}
\begin{eqnarray}
S_q &=& \:\frac{k_B}{1-q}
\left( \int (p_q)^q \:d\Omega-1 \right),
\label{eq:W1}
\end{eqnarray}
where $k_B$ denotes the Boltzmann constant,
$p_q$ the probability distribution function (PDF) 
and $\Omega$ the phase space element.
The Tsallis entropy is a one-parameter generalization of
that in the Boltzmann-Gibbs (BG) statistics, to which
Eq. (\ref{eq:W1}) reduces in the limit of $q=1.0$, 
\begin{eqnarray}
S_1 &=& S_{BG}= - k_B \:\int p_1 
\:\ln p_1\:d \Omega.
\label{eq:W2}
\end{eqnarray}
{\it Entropy additivity} is defined that an entropy $S$ is additive if 
for any two {\it probabilistically independent} systems A and B \cite{Tsallis09}:
\begin{eqnarray}
p_{q}(x_1, x_2) &=& p_q(x_1) \:p_q(x_2)
\hspace{1cm}\mbox{for $x_1 \in A,\; x_2 \in B$},
\label{eq:W3}
\end{eqnarray}
we have 
\begin{eqnarray}
S(A+B) &=& S(A)+(B).
\end{eqnarray} 
From the definition, the BG entropy is additive whereas
the Tsallis entropy is {\it nonadditive} \cite{Tsallis88,Tsallis98,Tsallis01,Tsallis04} 
because Eqs. (\ref{eq:W1}) and (\ref{eq:W3}) lead to
\begin{eqnarray}
S_{q}(A+B) &=& S_q(A)+S_q(B) +(1-q) S_q(A) \:S_q(B). 
\label{eq:W4}
\end{eqnarray}
For {\it probabilistically independent, identical} $N$-body systems,
an entropy $S(N)$ is additive if we have $S(N)=N S(1)$ for $\forall N$.

{\it Entropy extensivity} is a concept more subtle than the additivity 
in some sense \cite{Tsallis09}. 
An entropy $S(N)$ of a given $N$-body system is said to be {\it extensive} if
\begin{eqnarray}
0 < \lim_{N \rightarrow \infty} \;\frac{S(N)}{N} < \infty,
\end{eqnarray}
{\it i.e.,} if $S(N) \propto N$ for $N \gg 1$ \cite{Tsallis09}.
Additivity depends only on the specific mathematical connection 
between the macroscopic entropy functional and PDFs of the system.
Extensivity depends on this but also 
on the nature of the interaction and/or correlation of its elements. 
The distinction between additivity and extensivity has already been illustrated in simple
probabilistic systems (see Fig. 1 in Ref. \cite{Tsallis09}).
For independent (or short-ranged interaction) systems, 
the BG entropy is extensive 
while the Tsallis entropy is nonextensive.
In contrast, the BG entropy is nonextensive for systems
with long-ranged interactions. 
The Tsallis entropy may be extensive
for correlated systems such as the probabilistic system \cite{Tsallis05},
some fermionic system \cite{Caruso07},
and Gaussian PDFs with correlation \cite{Wilk07}.
The nonextensivity is realized in four different classes of systems
\cite{Tsallis04}:
(a) systems with long-range interactions, (b) small-scale systems
with fluctuations of temperatures or energy dissipations,
(c) systems coupled to finite-size thermal reservoir, and
(d) multi-fractal systems.
The Tsallis nonextensive statistics has been successfully applied to
a wide range of nonextensive systems including physics,
chemistry, astronomy, geophysics, biology, economics, and others \cite{Nonext}.

The specific heat and Tsallis entropy
in nonextensive systems such as ideal gases 
\cite{Plastino94}-\cite{Liyan08}
and harmonic oscillators \cite{Liyan08,Lenzi01,Lenzi02}
have been calculated mostly by using the $q$-average which is employed in
the un-normalized maximum-entropy method (MEM) \cite{Tsallis98}, 
normalized MEM \cite{Curado91}, 
and the optimum Lagrange multiplier (OLM) MEM \cite{Martinez00}.
This is because the MEM with the $q$-average has been believed 
to be more superior than the original MEM \cite{Tsallis88} with the
normal average \cite{Tsallis04,Abe02,Abe05,Abe06}.
Recently, however, it has been pointed out that
thermodynamical quantities obtained by the $q$-average 
are unstable for a small change of the PDF, 
whereas those obtained by the normal average 
are stable \cite{Abe08,Abe09,Abe09b}. Ref. \cite{Ven10} has shown that
expectation values may be self-consistently formulated
with the form of normal averages 
in the generalized variational perturbation approximation. 
In contrast, it has been claimed that
the Tsallis entropy and thermodynamical averages are robust
for the $q$-average \cite{Hanel09}.
The stability (robustness) of thermodynamical averages which 
is one of fundamental issues in the nonextensive
thermodynamics, is currently controversial \cite{Lutsko09}.
It is worthwhile to make calculations of  
thermodynamical quantities in $N$-body nonextensive systems
with the use of both the $q$- and normal averages.
Although the $N$ dependences of specific heats
in ideal gases and harmonic oscillators
have been calculated \cite{Plastino94}-\cite{Lenzi02},
those of the Tsallis entropy have not been investigated,
as far as we are aware of.
The purpose of the present paper is twofold: (1) to calculate the $N$ dependence
of thermodynamical quantities of nonextensive ideal gases and harmonic oscillators
and (2) to make a comparison between results with the $q$- and normal averages.
Such calculations are expected to provide some insight to the current 
controversy on the two average methods \cite{Abe08}-\cite{Lutsko09}.

One of alternative approaches to nonextensive systems is the superstatistics 
\cite{Wilk00}-\cite{Beck07}. Complex nonextensive systems are expected to 
temporary and spatially fluctuate. 
In the superstatistics, it is assumed that locally 
the equilibrium state is described by the Boltzmann-Gibbs statistics, 
and that their global properties may be expressed 
by a superposition of them over some intensive physical quantity,
{\it e.g.} the inverse temperature \cite{Wilk00}-\cite{Beck07}.
Many applications of the concept of the superstatistics have been 
pointed out (for a recent review, see \cite{Beck07}).
It is, however, not clear how to obtain the mixing probability 
distribution of fluctuating parameter from first principles.
This problem is currently controversial and
some attempts to this direction have been proposed
\cite{Tsallis03}-\cite{Straeten08}.
Previous studies in the superstatistics have paid attention mostly to
the stationary PDF, which generally has the non-Gaussian form.
However, thermodynamical quantities such as energy and entropy
have not been calculated in the superstatistics, 
which will be studied in this paper.

The paper is organized as follows:
In Sec. II, we briefly review the OLM-MEM 
with the $q$-average \cite{Martinez00}
and the original MEM with the normal average \cite{Tsallis88,Abe08}.
MEMs with the two average methods are applied to 
ideal gases and harmonic oscillators in Secs. III and IV, respectively.
In Sec. V, the fluctuation of the energy and
applications of the superstatistics \cite{Wilk00}-\cite{Beck07}
to $N$-body systems are studied.
Sec. VI is devoted to our conclusion.

\section{Maximum-entropy methods}
\subsection{OLM-MEM with the $q$-average}
We consider an independent $N$-body system 
whose hamiltonian is given by
\begin{eqnarray}
H &=& \sum_{i=1}^{N} H_i.
\label{eq:A1}
\end{eqnarray}
In order to obtain the $N$-variate density matrix $\rho_q$ 
from the Tsallis entropy with the OLM-MEM \cite{Martinez00},
we impose the constraints given by
\begin{eqnarray}
1&=& {\rm Tr} \:\rho_q,
\label{eq:A2} \\
U_q &=& [H]_q= \frac{{\rm Tr}\:\{ (\rho_q)^q  H \}}{c_q},
\label{eq:A3} 
\end{eqnarray}
with
\begin{eqnarray}
c_q &=& {\rm Tr} \: (\rho_q)^q,
\label{eq:A5}
\end{eqnarray}
where the bracket $[\cdot]_q$ denotes the $q$ average,
Tr the trace and $U_q$ the $q$-averaged energy. 
The OLM-MEM yields the density matrix given by \cite{Martinez00}
\begin{eqnarray}
\rho_q &=& \frac{1}{X_q} \:\left[1-(1-q)\gamma
\left( H - U_q  \right) \right]_+^{1/(1-q)},
\label{eq:A6}
\end{eqnarray}
with
\begin{eqnarray}
X_q&=& {\rm Tr} \:  \:\left[1-(1-q)\gamma
\left( H - U_q  \right) \right]_+^{1/(1-q)},
\label{eq:A7}
\end{eqnarray}
where $[x]_+={\rm max}(x,0)$ and $\gamma$ expresses a Lagrange multiplier.
Although $\beta$ is conventionally used for the relevant Lagrange
multiplier, we employ $\gamma$ to avoid a confusion
with the inverse of physical temperature ($T=1/k_B \beta$).
The Tsallis entropy defined by Eq. (\ref{eq:W1}) is expressed 
in terms of $c_q$ by
\begin{eqnarray}
S_q &=& \frac{c_q-1}{1-q},
\end{eqnarray}
where $c_q$ is given by Eq. (\ref{eq:A5}).

Next we relate the Lagrange multiplier $\gamma$
to the physical temperature $T$. 
It is an open problem how to define the physical temperature
in conformity with the zeroth law of thermodynamics 
in the nonextensive statistics \cite{Abe01f}-\cite{Abe06b}.
Among several schemes proposed for defining the physical temperature
\cite{Abe01f}-\cite{Abe06b}, the method of Ref. \cite{Abe01f} has been employed
in this study.
We assume that the physical temperature $T$ is given by \cite{Abe01f}
(for details, see Appendix A, related discussions being also given in Sec. VB)
\begin{eqnarray}
\frac{1}{T} &=& \frac{1}{c_q}
\frac{\partial S_q}{\partial U_q}= k_B\:\gamma,
\label{eq:A22}
\end{eqnarray}
which yields
\begin{eqnarray}
\gamma &=& \frac{1}{k_B \:T}=\beta.
\label{eq:A20}
\end{eqnarray}
We may rewrite Eq. (\ref{eq:A6}) as
\begin{eqnarray}
\rho_q
&=& \frac{1}{Z_{q} } \left[1-(1-q) 
\alpha \:H \right]_+^{1/(1-q)},
\label{eq:A8}
\end{eqnarray}
with
\begin{eqnarray}
Z_q&=& {\rm Tr}  \:\left[1-(1-q)\alpha
H \right]_+^{1/(1-q)},
\label{eq:A9}
\end{eqnarray}
\begin{eqnarray}
\alpha &=& \frac{\beta}{1+(1-q)\beta U_q}.
\label{eq:A10}
\end{eqnarray}
$U_q$ is expressed by
\begin{eqnarray}
U_q&=& \frac{1}{\nu_q Z_q}{\rm Tr}  \:\{\left[1-(1-q)\alpha
H \right]_+^{q/(1-q)} H\},
\label{eq:A9b}
\end{eqnarray}
with
\begin{eqnarray}
\nu_q &=& [1+(1-q)\beta U_q],
\end{eqnarray}
where we adopt the relation: $c_q=(X_q)^{1-q}=\nu_q (Z_q)^{1-q}$.

\subsection{Original MEM with the normal average}
In the original MEM with the normal average \cite{Tsallis88},
we impose the constraints given by
(tildes are attached to quantities relevant to the normal average)
\begin{eqnarray}
1 &=& {\rm Tr} \:\tilde{\rho}_q, 
\label{eq:A11}
\\
\tilde{U}_q &=& \langle H \rangle_q = {\rm Tr} \: \{\tilde{p}_q H\},
\label{eq:A12} 
\end{eqnarray}
where the bracket $\langle \cdot \rangle_q$ expresses 
the normal average and $\tilde{U_q}$ the normal-averaged energy.
The original MEM \cite{Tsallis88,Abe08} yields the density matrix give by
\begin{eqnarray}
\tilde{p}_q &=&\frac{1}{\tilde{X}_q} \left[1-(q-1)\tilde{\gamma}
\left( H - \tilde{U_q}  \right) \right]_+^{1/(q-1)},
\label{eq:A13}
\end{eqnarray}
with
\begin{eqnarray}
\tilde{X}_q &=& {\rm Tr} \: \left[1-(q-1) \tilde{\gamma}
\left( H - \tilde{U_q}  \right) \right]_+^{1/(q-1)},
\label{eq:A14}
\end{eqnarray}
where $\tilde{\gamma}$ denotes a Lagrange multiplier.
The Tsallis entropy is expressed by
\begin{eqnarray}
S_q &=& \frac{\tilde{c}_q-1}{1-q},
\end{eqnarray}
with
\begin{eqnarray}
\tilde{c}_q &=& {\rm Tr}\:(\tilde{\rho}_q)^q.
\end{eqnarray}

We assume that the physical temperature $T$ is given by \cite{Abe01f}
(for a detailed discussion, see Appendix A)
\begin{eqnarray}
\frac{1}{T} &=& \frac{1}{\tilde{c}_q}
\frac{\partial S_q}{\partial\tilde{U_q} }
=q \:k_B\:\tilde{\gamma},
\end{eqnarray}
which leads to
\begin{eqnarray}
\tilde{\gamma} &=& \frac{\beta}{q}.
\label{eq:A21}
\end{eqnarray}

Equation (\ref{eq:A13}) may be rewritten as
\begin{eqnarray}
\tilde{p}_q
&=& \frac{1}{\tilde{Z}_{q}} \left[1-\left(\frac{q-1}{q} \right)
\tilde{\alpha} \:H \right]_+^{1/(q-1)},
\label{eq:A15}
\end{eqnarray}
with
\begin{eqnarray}
\tilde{Z}_q &=& {\rm Tr}\:
\left[1-\left(\frac{q-1}{q} \right)
\tilde{\alpha} \:H \right]_+^{1/(q-1)},
\label{eq:A16}
\end{eqnarray}
\begin{eqnarray}
\tilde{\alpha} &=& \frac{ \beta/q }{1-(1/q-1) \beta \tilde{U_q} }.
\label{eq:A17}
\end{eqnarray}
$\tilde{U}_q$ is expressed by
\begin{eqnarray}
\tilde{U}_q &=& \frac{1}{\tilde{Z}_q}{\rm Tr}\:
\{\left[1-\left(\frac{q-1}{q} \right)
\tilde{\alpha} \:H \right]_+^{1/(q-1)}\:H\}.
\label{eq:A18}
\end{eqnarray}

Calculation methods discussed in this section
will be applied to classical ideal gases and harmonic oscillators in
Secs. III and IV, respectively.

\section{Ideal gases}
\subsection{$q$-average}
We consider hamiltonian for $N$-body $D$-dimensional ideal gases given by
\begin{eqnarray}
H = \sum_{i=1}^N\: H_i = \sum_{i=1}^N\: \frac{\vecp_i^2}{2m},
\label{eq:C1}
\end{eqnarray}
where $m$ stands for the mass and $\vecp_i$ momentum of ideal gases.
By using the exact method \cite{Prato95,Rajagopal98}, we obtain the partition function
$Z_q$ in Eq. (\ref{eq:A9}) (for details, see Appendix B1), 
\begin{eqnarray}
Z_{q} = \left\{ \begin{array}{ll}
\frac{V^N}{N! \:h^{DN}}\left[\frac{2 \pi\: m}{(q-1) \alpha} \right]^{D N/2}
\frac{\Gamma\left(\frac{1}{q-1}-\frac{D N}{2} \right)}
{\Gamma \left(\frac{1}{q-1} \right)}
\quad & \mbox{for $1 < q <3$}, \\
\frac{V^N}{N! \:h^{DN}}\left( \frac{2 \pi\: m}{\alpha} \right)^{D N/2}
\quad & \mbox{for $q=1$}, \\
\frac{V^N}{N! \:h^{DN}}\left[\frac{2 \pi m}{(1-q) \alpha} \right]^{D N/2}
\frac{\Gamma\left(\frac{1}{1-q}+1 \right)}
{\Gamma \left(\frac{1}{1-q}+\frac{D N}{2}+1 \right)}
\quad & \mbox{for $ q <1$}, \\
\end{array} \right.
\label{eq:C2}
\end{eqnarray}
where $h$ stands for the Planck constant and $V$ a volume of the system.
The average energy in Eq. (\ref{eq:A9b}) is given by (see Appendix B1) 
\begin{eqnarray}
U_q &=& \frac{D N}{[DN(1-q)+2] \alpha }.
\label{eq:C3}
\end{eqnarray}
From Eqs. (\ref{eq:A10}) and (\ref{eq:C3}),
$U_q$ and $\alpha$ are self-consistently determined as
\begin{eqnarray}
U_q &=& \frac{D N}{2 \beta}=\frac{k_B DNT}{2}, 
\label{eq:C4}\\
\alpha &=& \frac{\beta}{1+\frac{D N}{2} (1-q)}.
\label{eq:C5}
\end{eqnarray}
A conceivable $q$ value is given by
\begin{eqnarray}
0 < q < 1+\frac{2}{D N} \equiv q_U,
\label{eq:C8}
\end{eqnarray}
because $\rho_q$ given by Eq. (\ref{eq:A8}) has the probability properties with
$\alpha > 0$ and because the Tsallis entropy is stable for $q>0$ \cite{Abe02}.
Equation (\ref{eq:C4}) leads to the specific heat given by
\begin{eqnarray}
C_q &=& \frac{k_B D N }{2},
\label{eq:C10}
\end{eqnarray}
which agrees with that in the BG statistics. 
The specific heat of nonextensive ideal gases was discussed in Ref. \cite{Abe99}
by using the $q$-average with the un-normalized MEM \cite{Tsallis98},
where a possibility of the negative specific heat was pointed out. 
Our result of Eq. (\ref{eq:C10}) is
in agreement with previous studies using the OLM-MEM
\cite{Plastino94}-\cite{Abe01b}\cite{Liyan08}\cite{Lenzi01}.

The Tsallis entropy may be calculated by 
\begin{eqnarray}
S_q &=& k_B \:\left( \frac{c_q-1}{1-q} \right),
\label{eq:C11} 
\end{eqnarray}
with
\begin{eqnarray}
c_q &=& (X_q)^{1-q}=\nu_q (Z_{q})^{1-q},
\label{eq:C7b} \\
\nu_q &=& \frac{DN}{2}(1-q)+1.
\label{eq:C7} 
\end{eqnarray}
The BG entropy in the limit of $q=1.0$ is given by
\begin{eqnarray}
S_1 &=& N S_1^{(1)},
\label{eq:C12b}
\end{eqnarray}
with
\begin{eqnarray}
S_1^{(1)} &=&  k_B \left[
\left( \frac{D}{2} \right) \ln \left(\frac{2 \pi m}{h^2 \beta} \right) 
+  \ln \sigma  + \left( \frac{D}{2}+1 \right) \right]
\label{eq:C12}
\hspace{1cm}\mbox{($\sigma=V/N$)}.
\end{eqnarray}
By using the formula:
\begin{eqnarray}
\ln \left(\frac{\Gamma(z+a)}{\Gamma(z) z^a} \right)
\simeq -\frac{a(1-a)}{2 z}
\hspace{1cm}\mbox{for $\vert z \vert \rightarrow \infty $},
\end{eqnarray}
we obtain $S_q$ for $\vert q-1 \vert \ll 1.0$ given by
\begin{eqnarray}
S_q &=& N S_1^{(1)} -(q-1)\left[\frac{N^2}{2}
\left( S_1^{(1)} \right)^2 -\frac{D N}{4}
\right]+\cdot\cdot,
\label{eq:D12}
\end{eqnarray}
where the $(q-1)$ term includes the $O(N^2)$ 
contribution showing the nonadditivity.

The Tsallis entropy for large $N(1-q)$ is given by 
\begin{eqnarray}
S_q&\simeq & \left(\frac{k_B N}{2}\right) \: e^{(1-q)N S_1^{(1)}/k_B}
\hspace{1cm}\mbox{for $N (1-q) \gg 1$},
\label{eq:C13}
\end{eqnarray}
where we adopt the relation:
$\ln \Gamma(z) \simeq (z-1/2) \ln z-z+(1/2)\ln 2\pi+ \cdot\cdot$ 
for $\vert z \vert \rightarrow \infty$.


\subsection{Normal average}
For ideal gases whose hamiltonian is given by Eq. (\ref{eq:C1}),
we obtain (for details, see Appendix B2)
\begin{eqnarray}
\tilde{Z}_{q} = \left\{ \begin{array}{ll}
\frac{V^N}{N!\:h^{D N}} \left[\frac{2 \pi\: m}{(q-1) \tilde{\alpha} } \right]^{D N/2}
\frac{\Gamma\left(\frac{1}{q-1}+1 \right)}
{\Gamma \left(\frac{1}{q-1}+\frac{D N}{2}+1 \right)}
\quad & \mbox{for $1 < q <3$}, \\
\frac{V^N}{N!\:h^{D N}} \left( \frac{2 \pi\: m}{\tilde{\alpha} } \right)^{D N/2}
\quad & \mbox{for $q=1$}, \\
\frac{V^N}{N!\:h^{D N}} \left[\frac{2 \pi m}{(1-q) \tilde{\alpha} }  \right]^{D N/2}
\frac{\Gamma\left(\frac{1}{1-q}-\frac{D N}{2} \right)}
{\Gamma \left(\frac{1}{1-q} \right)}
\quad & \mbox{for $ q <1$}. \\
\end{array} \right.
\label{eq:D1}
\end{eqnarray}
$\tilde{U_q}$ in Eq. (\ref{eq:A18}) may be expressed in terms of $\tilde{\alpha} $ 
(see Appendix B2),
\begin{eqnarray}
\tilde{U_q} &=& \frac{D N}{[DN(q-1)+2q ] \tilde{\alpha} }.
\label{eq:D2}
\end{eqnarray}
From Eqs. (\ref{eq:A17}) and (\ref{eq:D2}),
$\tilde{U_q}$ and $\tilde{\alpha}$ are self-consistently determined as
\begin{eqnarray}
\tilde{U_q} &=& \frac{D N}{2 \beta}=\frac{k_B DNT}{2}, 
\label{eq:D3}
\\
\tilde{\alpha} &=& \frac{\beta }{q+\frac{DN}{2} (q-1)}.
\label{eq:D4}
\end{eqnarray}
A conceivable $q$ value is given by
\begin{eqnarray}
q > 1-\frac{1}{\left( \frac{D N}{2} +1\right)} \equiv q_L =\frac{1}{q_U},
\label{eq:D7}
\end{eqnarray}
where 
$q_U$ is given by Eq. (\ref{eq:C8}).
Equation (\ref{eq:D3}) yields specific heat given by
\begin{eqnarray}
\tilde{C}_q &=& \frac{k_B D N}{2},
\label{eq:D10} 
\end{eqnarray}
which is the same as that of the BG statistics.

The Tsallis entropy is expressed by
\begin{eqnarray}
\tilde{S}_q &=& k_B \:\left(\frac{\tilde{c}_q-1}{1-q} \right),
\label{eq:D11}
\end{eqnarray}
with
\begin{eqnarray}
\tilde{c}_q &=& 
(\tilde{X}_q)^{1-q}
= \frac{(\tilde{Z}_{q})^{1-q}}{\tilde{\nu}_q},
\label{eq:D6c} \\
\tilde{\nu}_q &=& \frac{DN}{2}\left(1-\frac{1}{q} \right)+1
= \nu_{1/q},
\label{eq:D6b}
\end{eqnarray}
where $\nu_q$ is given by Eq. (\ref{eq:C7}).

For $\vert q-1 \vert \ll 1.0$, we obtain $\tilde{S}_q$ given by
\begin{eqnarray}
\tilde{S}_q 
&=& N S_1^{(1)} -(q-1)\left[\frac{N^2}{2}
\left( S_1^{(1)} \right)^2 + \frac{D N}{4}
\right]+\cdot\cdot,
\label{eq:D14}
\end{eqnarray}
where $S_1^{(1)}$ is given by Eq. (\ref{eq:C12}).
Equation (\ref{eq:D14}) is the same as Eq. (\ref{eq:D12})
except for the $O(DN)$ term in the bracket. 

For large $N(q-1)$, $\tilde{S}_q$ is given by 
\begin{eqnarray}
\tilde{S}_q &\simeq& \frac{k_B}{q-1} \left[1 
-\frac{2q\:e^{-(q-1)N S_1^{(1)}/k_B}}{N(q-1)}  \right] 
\hspace{0.5cm}\mbox{for $N(q-1) \gg 1$},
\label{eq:D13}
\end{eqnarray}
which is quite different from $S_q$ in Eq. (\ref{eq:C13}) 
obtained by the $q$-average.

\subsection{Model calculations}

We will present some numerical calculations of the Tsallis entropy of ideal gases 
with $D=1$. Equations (\ref{eq:C8}) and (\ref{eq:D7}) show that
the conceivable $q$ values for $q$- and normal averages
are $0 < q < q_U$ and $q > q_L$, respectively, where
$q_U= 1+2/DN$ and $q_L = 1-2/(DN+2)$ (see the inset of Fig. \ref{fig1}). 
$q_L$ and $q_U$ approach unity for $N \rightarrow \infty$.
Although the $N$ dependence of the Tsallis entropy in the $q$-average
is rather different from that in the normal average, 
as Eqs. (\ref{eq:C13}) and (\ref{eq:D13}) show,
both results are in fairly good agreement for $q_L < q < q_U$ where both the 
$q$- and normal averages are valid.
Dashed and solid curves in Fig. \ref{fig1} show the entropies 
calculated by the $q$- and normal averages, respectively, 
for ideal gases with $N=100$ and $T/T_0=1.0$
where $T_0=h^2/k_B m$ ($q_L=0.9804$ and $q_U=1.020$).
The difference between the two entropies  
is small for $q_L < q < q_U$ as shown by the bold curve
in Fig. \ref{fig1} except for $q \lesssim q_L$ where $\tilde{S}_q$ 
is divergently increased. 

Figure \ref{fig2} shows the three-dimensional plot of
$S_q/k_B N$ calculated by the $q$-average as functions of $q$ and $N$ for $T/T_0=1.0$.
With increasing $N$, $S_q$ is considerably increased, as Eq. (\ref{eq:C13})
shows: note the logarithmic scale in the ordinate.

The temperature dependence of $S_q$ calculated by the $q$-average is shown 
in Figs. \ref{fig3} and \ref{fig4}: in the former
$S_q/k_B N$ for $N=100$ is plotted as functions of $T$ and $q$,
while in the latter $S_q/k_B N$ for $q=0.8$ 
is plotted as functions of $T$ and $N$.
With increasing the temperature, the Tsallis entropy is increased,
as expected. This temperature dependence becomes more significant
for larger $N$.

Figure \ref{fig5} shows $\tilde{S}_q/k_B N$ calculated by the normal average 
for $T/T_0=1.0$ as functions of $q$ and $N$.
The magnitude of $\tilde{S}_q/k_B N$ is significantly decreased with
increasing $N$ and/or $q$, as Eq. (\ref{eq:D13}) shows. 
This behavior is quite different
from that obtained by the $q$-average shown in Fig. \ref{fig2}.

Figure \ref{fig6} shows $\tilde{S}_q/k_B N$ calculated by the normal average 
as functions of $T$ and $q$,
whereas Fig. \ref{fig7} shows $\tilde{S}_q/k_B N$ as functions of $T$ and $N$.
With increasing the temperature, $\tilde{S}_q$ is increased.
The temperature dependence of $\tilde{S}_q$ becomes weaker
for larger $N$ and/or larger $q$.

\section{Harmonic oscillators}

\subsubsection{$q$-average}
We consider hamiltonian for $N$-body $D$-dimensional harmonic oscillators given by
\begin{eqnarray}
H = \sum_{i=1}^N\: H_i 
= \sum_{i=1}^N\: \left( \frac{\vecp_i^2}{2m}+\frac{m \omega^2 \vecr_i^2}{2} \right),
\label{eq:E1}
\end{eqnarray}
where $m$, $\omega$, $\vecr_i$ and $\vecp_i$ express the mass,
oscillator frequency, position and momentum, respectively.
From Eq. (\ref{eq:A9}), we obtain (see Appendix B1)
\begin{eqnarray}
Z_{q} = \left\{ \begin{array}{ll}
\frac{1}{h^{D N}} \left[\frac{2 \pi}{(q-1) \omega \alpha} \right]^{D N}
\frac{\Gamma\left(\frac{1}{q-1}-D N \right)}
{\Gamma \left(\frac{1}{q-1} \right)}
\quad & \mbox{for $1 < q <3$}, \\
\frac{1}{h^{D N}} \left( \frac{2 \pi}{\omega \alpha} \right)^{D N}
\quad & \mbox{for $q=1$}, \\
\frac{1}{h^{D N}} \left[ \frac{2 \pi }{(1-q) \omega \alpha} \right]^{D N}
\frac{\Gamma\left(\frac{1}{1-q}+1 \right)}
{\Gamma \left(\frac{1}{1-q}+1+D N \right)}
\quad & \mbox{for $ q <1$}. \\
\end{array} \right.
\label{eq:E2}
\end{eqnarray}
$U_q$ in Eq. (\ref{eq:A9b}) is expressed in terms of $\alpha$ (see Appendix B1),
\begin{eqnarray}
U_q &=& \frac{D N}{[DN(1-q)+1] \alpha }.
\label{eq:E3}
\end{eqnarray}
From Eqs. (\ref{eq:A10}) and (\ref{eq:E3}),
$U_q$ and $\alpha$ are self-consistently determined as
\begin{eqnarray}
U_q &=& \frac{D N}{\beta}=k_B \:DNT, 
\label{eq:E4}\\
\alpha &=& \frac{\beta}{1+DN(1-q)}
\hspace{1cm}\mbox{for $0 < q < 1+\frac{1}{DN} \equiv q_U$}.
\label{eq:E5}
\end{eqnarray}
Equation (\ref{eq:E4}) yields the specific heat given by
\begin{eqnarray}
C_q &=& k_B D N. 
\label{eq:E9}
\end{eqnarray}
The specific heat of harmonic oscillators 
was discussed in Ref. \cite{Lenzi01}
with the use of un-normalized \cite{Curado91} 
and normalized MEMs \cite{Tsallis98}.
Their expressions for the internal energy $U_q$ have rather 
complicated $q$ dependence [see Eqs. (12) and (22) of Ref. \cite{Lenzi01}].
Equation (\ref{eq:E9}) expresses the Dulong-Petit specific heat:
$C_{BG}=k_B D N$ in the BG statistics,  
which is in agreement with Ref. \cite{Liyan08}.

The Tsallis entropy is given by Eq. (\ref{eq:C11}) with
\begin{eqnarray}
c_q &=& \nu_q (Z_{q})^{1-q},
\label{eq:E7} \\
\nu_q &=& DN (1-q)+1.
\label{eq:E7b}
\end{eqnarray}
The BG entropy in the limit of $q=1.0$ is given by 
\begin{eqnarray}
S_1=N S_1^{(1)}, 
\label{eq:E10b}
\end{eqnarray}
with
\begin{eqnarray}
S_1^{(1)} &=& k_B D \left[\ln\left(\frac{2 \pi}{h \omega \beta} \right) 
+1 \right].
\label{eq:E10}
\end{eqnarray}

\subsubsection{Normal average}
For harmonic oscillators under consideration,
we obtain (see Appendix B2)
\begin{eqnarray}
\tilde{Z}_{q} = \left\{ \begin{array}{ll}
\frac{1}{h^{D N}}\left[\frac{2 \pi}{(q-1) \omega \tilde{\alpha} } \right]^{D N}
\frac{\Gamma\left(\frac{1}{q-1}+1\right)}
{\Gamma \left(\frac{1}{q-1}+D N+1 \right)}
\quad & \mbox{for $1 < q <3$}, \\
\frac{1}{h^{D N}}\left( \frac{2 \pi}{\omega  \tilde{\alpha} } \right)^{D N}
\quad & \mbox{for $q=1$}, \\
\frac{1}{h^{D N}}\left[\frac{2 \pi }{(1-q) \omega  \tilde{\alpha} } \right]^{D N}
\frac{\Gamma\left(\frac{1}{1-q}-D N \right)}
{\Gamma \left(\frac{1}{1-q} \right)}
\quad & \mbox{for $ q <1$}. \\
\end{array} \right.
\label{eq:F1}
\end{eqnarray}
We may express $\tilde{U_q}$ given by Eq. (\ref{eq:A18}) 
in terms of $ \tilde{\alpha} $ as given by (see Appendix B2)
\begin{eqnarray}
\tilde{U_q} &=& \frac{D N}{[DN (q-1)+q] \tilde{\alpha}  }.
\label{eq:F2}
\end{eqnarray}
From Eqs. (\ref{eq:A17}) and (\ref{eq:F2}),
$\tilde{U_q}$ and $ \tilde{\alpha} $ are self-consistently determined as
\begin{eqnarray}
\tilde{U_q} &=& \frac{DN}{\beta}=k_B\:DNT, 
\label{eq:F3}\\
\tilde{\alpha}  &=& \frac{\beta }{q+DN(q-1)}
\hspace{1cm}\mbox{for $q >1-\frac{1}{(DN+1)} \equiv q_L =1/q_U$},
\label{eq:F4}
\end{eqnarray}
where 
$q_U$ is given by Eq. (\ref{eq:E5}).
We obtain the specific heat given by
\begin{eqnarray}
\tilde{C}_q &=& k_B D N.
\label{eq:F9}
\end{eqnarray}
Calculated specific heat is the same as the Boltzmann-Gibbs result
of $C_{BG}= k_B D N$.

The Tsallis entropy may be calculated by Eq. (\ref{eq:D11}) with
\begin{eqnarray}
\tilde{c}_{q} &=& 
\frac{(\tilde{Z}_{q})^{1-q}}{\tilde{\nu}_q},
\label{eq:F6} \\
\tilde{\nu}_q &=& DN \left(1-\frac{1}{q}\right)+1=\nu_{1/q},
\label{eq:F6b}
\end{eqnarray}
where $\nu_q$ is given by Eq. (\ref{eq:E7b}).

The properties of the specific heat and Tsallis entropies of harmonic oscillators
are essentially the same as those of ideal gases
if we read $DN/2 \rightarrow DN$ and $m \rightarrow 1/\omega$.

\section{Discussion}

\subsection{Fluctuation of the energy}
It is interesting to examine the fluctuation of energy in nonextensive
systems, which has been studied in Refs. \cite{Abe99b,Liyan08,Feng10}.
The $q$-average of $H^2$ is given by (see Appendix B1)
\begin{eqnarray}
[H^2]_q &=& \frac{\eta DN \left(\eta DN +1 \right)[1+\eta DN(1-q)]}
{\beta^2 [1+(\eta DN+1)(1-q)]},
\label{eq:V1}
\end{eqnarray}
where $\eta=1/2$ and $\eta=1$ for ideal gases and harmonic oscillators,
respectively.
The relative fluctuation of the energy is given by
\begin{eqnarray}
\frac{\Delta U_q}{U_q} &\equiv& \frac{\sqrt{[H^2-U_q^2]_q}}{U_q}
=\frac{1}{ \sqrt{\eta DN [1+(\eta DN+1)(1-q)]} },
\label{eq:V2}\\
&\propto & \left\{ \begin{array}{ll}
\frac{1}{\sqrt{DN}}
\quad & \mbox{for $q = 1.0$}, \\
\frac{1}{DN}
\quad & \mbox{for $DN (1-q) \gg  1$}. 
\end{array} \right.
\label{eq:V3}
\end{eqnarray}
Equation (\ref{eq:V2}) agrees with the previous results 
with the OLM-MEM \cite{Abe99b,Liyan08,Feng10}.
Equation (\ref{eq:V3}) shows that when the nonextensivity exists, the relative
fluctuation is proportional to $1/N$ for large $N$ 
instead of $1/\sqrt{N}$ in the BG statistics.

The normal average of $H^2$ is given by (see Appendix B2)
\begin{eqnarray}
\langle H^2 \rangle_q &=& \frac{\eta DN \left( \eta DN+1 \right)[1+(\eta DN+1)(q-1)]}
{\beta^2 [1+(\eta DN+2)(q-1)]}, 
\label{eq:V4}
\end{eqnarray}
which leads to the relative fluctuation of the energy given by
\begin{eqnarray}
\frac{\Delta \tilde{U}_q}{\tilde{U}_q} 
&\equiv & \frac{\sqrt{\langle H^2-\tilde{U}_q^2\rangle_q}}{\tilde{U}_q}
= \frac{1}{\sqrt{\eta DN \left[1+(\eta DN +1) \left(1-1/q \right)\right]}  },
\label{eq:V5}\\
&\propto & \left\{ \begin{array}{ll}
\frac{1}{\sqrt{DN} }
\quad & \mbox{for $q = 1.0$}, \\
\frac{1}{DN}
\quad & \mbox{for $DN(q-1) \gg 1$}. 
\end{array} \right.
\label{eq:V6}
\end{eqnarray}
Equation (\ref{eq:V5}) has the similar $DN$ dependence to Eq. (\ref{eq:V2}).
We note that Eqs. (\ref{eq:V2}) and (\ref{eq:V5}) have the reciprocal symmetry:
$q \leftrightarrow 1/q$. 

Figure \ref{fig8} shows the 3d plot of $\Delta U_q/U_q$
as functions of $q$ and $(\eta DN)$ calculated by the $q$ average.
The $DN$ dependence of $\Delta U_q/U_q$ for large $N$ is changed from $1/\sqrt{DN}$
to $1/DN$ when the nonextensivity is introduced as Eq. (\ref{eq:V3}) shows.
We notice also that the magnitude of $\Delta U_q/U_q$ is much decreased by the nonextensivity.
The $DN$ dependence of the relative fluctuation of energy
in the normal average is given in Fig. \ref{fig8} if we read
$q \rightarrow 1/q$.

\subsection{Superstatistics}
\subsubsection{Probability distribution function}
In Sec. II, we have derived the PDF by using the MEMs.
It is instructive to obtain the PDF in the superstatistics
\cite{Wilk00}-\cite{Beck07}.
The PDF of $D$-dimensional $N$-body ideal gases in the locally equilibrium state 
with a given inverse temperature $\beta$ is expressed by
\begin{eqnarray}
\Pi^{(N)}(\vecp;\beta) &=& \frac{N!}{V^N} 
\left( \frac{h^2 \beta}{2 \pi m}\right)^{D N/2}
\:e^{-\beta \sum_{i=1}^{N} \vecp_i^2/2m},
\label{eq:G1}
\end{eqnarray}
where $\vecp=\{ \vecp_i\}$ ($i=1$ to $N$).
We assume that the distribution of $\beta$ is given by 
the $\chi^2$-distribution with the rank $n$ \cite{Wilk00,Beck01},
\begin{eqnarray}
f(\beta) &=& \frac{1}{\Gamma(n/2)}\left( \frac{n}{2 \bar{\beta} }\right)^{n/2}
\beta^{n/2-1}\:e^{-n \beta/2 \bar{\beta}},
\label{eq:G2}
\end{eqnarray}
where $\bar{\beta}$ stands for the average of $\beta$ over $f(\beta)$:
$\langle \beta \rangle_{\beta}=\bar{\beta}$.
The PDF averaged over $f(\beta)$ is given by
\begin{eqnarray}
p^{(N)}(\vecp) &=& \int_0^{\infty} \Pi^{(N)}(\vecp;\beta)f(\beta)\:d \beta,\\
&=& \frac{1}{Z} \left[1 +\left(\frac{2 \bar{\beta}}{n} \right)
\sum_{i=1}^N \frac{\vecp_i^2}{2m} \right]_+^{-(D N+n)/2},
\label{eq:G3}
\end{eqnarray}
with
\begin{eqnarray}
Z &=& \frac{V^N}{N!} \left(\frac{\pi m n}{h^2 \bar{\beta}} \right)^{DN/2}
\frac{\Gamma(n/2)}{\Gamma(DN/2+n/2)}.
\label{eq:G6}
\end{eqnarray}

The PDF of harmonic oscillators in the local equilibrium state
with a given inverse temperature $\beta$ is expressed by
\begin{eqnarray}
\Pi^{(N)}(\vecp,\vecx;\beta) &=& \left( \frac{h \omega \beta}{2 \pi}\right)^{D N}
\:e^{-\beta \sum_{i=1}^{N}( \vecp_i^2/2m+ m \omega^2 \vecr_i^2/2)}.
\label{eq:H1}
\end{eqnarray}
The PDF averaged over $f(\beta)$ is given by
\begin{eqnarray}
p^{(N)}(\vecp,\vecx) 
&=& \int_0^{\infty} \Pi^{(N)}(\vecp,\vecx;\beta)f(\beta)\:d \beta, \\
&=& \frac{1}{Z}\left[1 +\left(\frac{2 \bar{\beta}}{n} \right)
\sum_{i=1}^N \left( \frac{\vecp_i^2}{2m}
+ \frac{m \omega^2 \vecr_i^2}{2} \right)\right]_+^{-(D N+n/2)}.
\label{eq:H2}
\end{eqnarray}
with
\begin{eqnarray}
Z &=& \left(\frac{\pi n}{h \omega \bar{\beta}}  \right)^{DN}
\frac{\Gamma(n/2)}{\Gamma(DN+n/2)}.
\label{eq:G7}
\end{eqnarray}

It is easy to see that
PDFs given by Eqs. (\ref{eq:G3}) and  (\ref{eq:H2}) have the same structure 
as those given by Eqs. (\ref{eq:A8}) and (\ref{eq:A15}) derived by the MEMs.
For example, Eq. (\ref{eq:G3}) for ideal gases may be rewritten as
\begin{eqnarray}
p^{(N)}(\vecp) &\propto& \left[1 -(1-q)\left(\frac{DN+n}{n}\right) \bar{\beta} 
\sum_{i=1}^N \frac{\vecp_i^2}{2m} \right]_+^{1/(1-q)},
\label{eq:G9}
\end{eqnarray}
with
\begin{eqnarray}
q &=& 1 + \frac{2}{DN+n},
\end{eqnarray}
or equivalently
\begin{eqnarray}
p^{(N)}(\vecp) &\propto& 
\left[1 -\left(\frac{\tilde{q}-1}{\tilde{q}}\right) 
\left(\frac{DN+n-2}{n}\right) \bar{\beta} 
\sum_{i=1}^N \frac{\vecp_i^2}{2m} \right]_+^{1/(\tilde{q}-1)},
\label{eq:G10}
\end{eqnarray}
with
\begin{eqnarray}
\tilde{q} &=& 1 - \frac{2}{DN+n}.
\end{eqnarray}
Equations (\ref{eq:G9}) and (\ref{eq:G10}) are similar to
Eqs. (\ref{eq:A8}) and (\ref{eq:A15}), respectively, although
the ranges of conceivable $q$ values are not necessarily the same.

\subsubsection{Energy and entropy}

The $\chi^2$-distribution in Eq. (\ref{eq:G2}) is given {\it ad hoc} 
\cite{Wilk00}-\cite{Beck07}.
Several first-principles approaches have been proposed
to determine the optimum distribution of $\beta$ in the superstatistics
\cite{Tsallis03}-\cite{Straeten08}.
Ref.\cite{Straeten08} considered the local energy $E(\beta)=\left< H \right>_{\Pi}$ 
and entropy $S(\beta)=- \left< \ln \Pi \right>_{\Pi}$ for local equilibrium
states with an inverse temperature $\beta$, and then
obtained the energy and entropy of the system given by
$E=\langle E(\beta)\rangle_{\beta}$ and $S=\langle S(\beta)\rangle_{\beta}$, 
where $\langle \cdot \rangle_{\Pi}$ and $\langle \cdot \rangle_{\beta}$
express averages over $\Pi(\beta)$ and $f(\beta)$, respectively.
Equations (\ref{eq:C4}) and (\ref{eq:C12}) for ideal gases lead to
\begin{eqnarray}
E(\beta) &=& \frac{DN}{2 \beta}, \\
S(\beta) &=&  k_B N \left[\left( \frac{D}{2} \right) \ln \left(\frac{2 \pi m}{h^2 \beta} \right) 
+  \ln \sigma  + \left( \frac{D}{2}+1 \right)
\right],
\label{eq:G8}
\end{eqnarray}
from which we obtain
\begin{eqnarray}
E &=&\frac{DN}{2 \bar{\beta}}\left( \frac{ n}{n-2}\right),
\label{eq:G4}\\
S &=&  k_B N \left[\left( \frac{D}{2} \right) \ln \left(\frac{2 \pi m}{h^2 \bar{\beta}} \right) 
+  \ln \sigma  + \left( \frac{D}{2}+1 \right)
\right]
- \frac{k_B DN}{2} \left[ \psi\left(\frac{n}{2}\right)-\ln\left(\frac{n}{2} \right) \right],
\label{eq:G5} 
\end{eqnarray}
$\psi(x)$ standing for the poli-gamma function.
Similarly from Eqs. (\ref{eq:E4}) and (\ref{eq:E10}) for harmonic oscillators, 
we obtain
\begin{eqnarray}
E(\beta) &=& \frac{DN}{\beta}, \\
S(\beta) &=& k_B D N \left[\ln\left(\frac{2 \pi}{h \omega \beta} \right) 
+1 \right],
\label{eq:H7}
\end{eqnarray}
and then
\begin{eqnarray}
E &=& \frac{DN}{\bar{\beta}}\left( \frac{ n}{n-2}\right), 
\label{eq:H4}\\
S &=& k_B D N \left[\ln\left(\frac{2 \pi}{h \omega \bar{\beta}} \right) 
+1 \right]
-k_B DN \left[ \psi\left(\frac{n}{2}\right)
-\ln\left(\frac{n}{2} \right) \right].
\label{eq:H5}
\end{eqnarray}

We note that results given by Eqs. (\ref{eq:G4}), (\ref{eq:G5}),
(\ref{eq:H4}) and (\ref{eq:H5}) are different from 
their counterparts obtained by the MEMs.
The similar result is obtained in the maximum-entropy approach to the superstatistics
in Ref. \cite{Abe07} based on the conditional entropy.
When we employ, in place of the gamma distribution of Eq. (\ref{eq:G2}),
the inverse-gamma distribution proposed in Ref. \cite{Touch02},
\begin{eqnarray}
f_T(\beta) &=& \frac{\bar{\beta}}{\Gamma(n/2)}
\left( \frac{n \bar{\beta}}{2 }\right)^{n/2}
\beta^{-n/2-2}\:e^{-n \bar{\beta}/2 \beta},
\label{eq:H6}
\end{eqnarray}
we obtain for ideal gases, 
\begin{eqnarray}
E &=& \frac{D N }{2 \bar{\beta}}
\left(\frac{n+2}{ n} \right), \\
S &=&k_B N \left[
\left( \frac{D}{2} \right) \ln \left(\frac{2 \pi m}{h^2 \bar{\beta}} \right) 
+  \ln \sigma  + \left( \frac{D}{2}+1 \right) \right]
-\frac{k_B DN}{2} \left[ \psi\left(\frac{n}{2}+1 \right)
-\ln\left(\frac{n}{2} \right) \right].
\label{eq:H8}
\end{eqnarray}
We note that entropies in Eqs. (\ref{eq:G5}), (\ref{eq:H5}) and (\ref{eq:H8}) 
derived by the maximum-entropy approaches to
the superstatistics \cite{Abe07,Straeten08} are additive,
which is in contrast with the nonadditive Tsallis entropy defined by Eq. (\ref{eq:W1}).
This result is not modified even when we adopt any distribution
function of $\beta$, because $S(\beta)$ is given by
$S(\beta)=k_B N h(\beta)$ [see Eqs. (\ref{eq:G8}) and (\ref{eq:H7})] 
and then $S=k_B N \langle h(\beta) \rangle_{\beta}$,
where $h(\beta)$ depends on $\beta$ but on $N$.
The superstatistics \cite{Wilk00,Beck01} was introduced to account 
for the non-Gaussian distribution which is well described by 
the $q$-Gaussian ($q$-exponential) form
originating from the nonadditive Tsallis entropy \cite{Tsallis88}.
We then naturally expect that the entropy obtained by the superstatistics  
is also nonadditive. It is, however, not the case
in the first-principles approaches to superstatistics \cite{Abe07,Straeten08}
as mentioned above.

Alternatively, we may calculate the energy and entropy by using
the partition function $Z$ given by Eqs. (\ref{eq:G6}) or (\ref{eq:G7}). 
For ideal gases, we obtain
\begin{eqnarray}
E' &=& - \frac{\partial Z}{\partial \bar{\beta}}
= \frac{DN}{2 \bar{\beta}}, \\
S' &=& k_B (\ln Z + \bar{\beta} E') \nonumber \\
&=& k_B N \left[
\left( \frac{D}{2} \right) \ln \left(\frac{2 \pi m}{h^2 \bar{\beta}} \right) 
+  \ln \sigma  + \left( \frac{D}{2}+1 \right) \right] \nonumber \\
&+& k_B \left[\frac{DN}{2} \ln\left(\frac{n}{2} \right) 
+ \ln \Gamma\left(\frac{n}{2} \right)
- \ln \Gamma\left(\frac{DN}{2}+\frac{n}{2} \right) \right].
\end{eqnarray}
$E'$ agrees with Eq. (\ref{eq:C4}) or (\ref{eq:D3}) if we read $\bar{\beta}=\beta$.
Although $S'$ is nonadditive, 
its $N$ dependence is different from $S_q$ derived by the MEM in Sec. III.

\subsection{The physical temperature}
As mentioned in the introduction, there are several proposals
in relating the Lagrange multiplier to the physical temperature
even within the OLM-MEM \cite{Abe01f}-\cite{Abe06b}.
In this paper, we have adopted the scheme proposed in Ref. \cite{Abe01f}.
One of the alternatives is to define the physical temperature
by $1/T=\partial S_q/\partial U_q$ \cite{Tsallis01}, 
yielding for ideal gases (with the OLM-MEM),
\begin{eqnarray}
\frac{1}{T} &=& \frac{\partial S_q}{\partial U_q}
= k_B \gamma  \:c_q,
\label{eq:J1}
\end{eqnarray}
which is different from Eq. (\ref{eq:A22}) by a factor of $c_q$.
Then the Lagrange multiplier is determined as
\begin{eqnarray}
\gamma &=& \frac{1}{k_B T \:c_q},
\label{eq:J2}
\end{eqnarray}
and the energy is given by
\begin{eqnarray}
U_q &=& \frac{D N k_B \:T \: c_q}{2}.
\label{eq:J3}
\end{eqnarray}
The expression for the specific heat becomes complex because
$c_q$ in Eq. (\ref{eq:J3}) is temperature dependent:
$Z_q$ in $c_q$ has the temperature dependence through $\gamma$. 

\section{Concluding remarks}

\begin{table}
\begin{center}
\caption{Specific heat $C_q$ and the Tsallis entropy $S_q$
of ideal gases 
[$S_1^{(1)}$ is given by Eqs. (\ref{eq:C12}),
$q_U=1+2/DN$ and $q_L = 1-1/(DN/2 +1)=1/q_U$.
In the superstatistics \cite{Abe07,Straeten08}, the $\chi^2$-distribution with rank $n$
is employed: $G(n)=\psi(n/2)-\ln(n/2) \propto 1/n$ for $n \gg 1$.
{\it NA}: not applicable]
}
\renewcommand{\arraystretch}{1.5}
\begin{tabular}{|c||c|c|c|c|} \hline
method
& range & \;\;$C_q$ \;\; &$S_q$ (for $\vert q-1 \vert \ll 1.0$) 
& \;\;$S_q$ (for $N \vert q-1 \vert \gg 1$) 
\\ \hline \hline
$q$-average & $0< q < q_U$
& $\frac{k_B D N}{2}$
& $N S_1^{(1)} - (q-1)\left[\frac{N^2}{2}\left(S_1^{(1)} \right)^2-\frac{DN}{4} \right]$
& $  \frac{k_B N}{2}\:e^{(1-q)NS_1^{(1)}/k_B}$  
\\ \hline
normal average & $q_L < q < \infty$
& $\frac{k_B D N}{2}$ 
& $N S_1^{(1)} - (q-1)\left[\frac{N^2}{2}\left(S_1^{(1)} \right)^2+\frac{DN}{4} \right]$
& $\frac{k_B}{(q-1)}$
\\ \hline
Superstatistics & $n > 2$
& $\frac{k_B D N n}{2(n-2)}$ 
& {\it NA} 
& $N \left[S_1^{(1)}-\frac{k_B D}{2}G(n) \right]$ 
\\ \hline
\end{tabular}
\end{center}
\end{table}

Specific heats and entropies in nonextensive ideal gases and harmonic oscillators
have been calculated by using the MEMs with the $q$- and normal averages
and by the superstatistics. Obtained results of ideal gases are
summarized in Table 1: those of harmonic oscillators
are similarly given if we read $DN/2 \rightarrow DN$. 
Our calculations have shown the followings: 

\noindent
(i) The MEMs with the $q$- and normal averages are valid 
for $0 < q < q_U$ and $q_L < q < \infty$, respectively,
where $q_U=1+1/\eta DN$ and $q_L=1-1/(\eta DN+1)$,

\noindent
(ii) Energy and specific heat obtained by the $q$ average coincide
with those derived by the normal average and in the BG statistics,

\noindent
(iii) the Tsallis entropy obtained by the $q$- and normal averages
are in fairly good agreement for $q_L < q < q_U$,

\noindent
(iv) the Tsallis entropy in the $q$-average shows
the exponential $N$ dependence: $S_q/N \sim e^{(1-q)N S_1^{(1)}}$
for $N(1-q) \gg 1$,
which is quite different from that in the normal average:
$\tilde{S}_q/N \sim 1/(q-1)N$ for $N(q-1) \gg 1$, and

\noindent
(v) the entropy in first-principles approaches to the superstatistics of Refs.
\cite{Abe07,Straeten08} is additive ($S^{(N)} = N S^{(1)}$) 
while the Tsallis entropy is nonadditive.

\noindent
Items (i), (iii) and (iv) are consistent with results obtained with the $q$-Gaussian PDF
\cite{Hasegawa10}.
The item (ii) shows that the rule of energy equipartition holds in both BG and
nonextensive statistics.
The item (iii) implies that the calculated properties of the entropy
considerably depend on the adopted average methods \cite{Hasegawa10}.
It would be necessary to further examine $q$- and normal averages
from various viewpoints \cite{Abe09b}-\cite{Lutsko09} and
to develop a first-principles superstatistical method
which yields an entropy in conformity with the Tsallis one.

\begin{acknowledgments}
This work is partly supported by
a Grant-in-Aid for Scientific Research from 
Ministry of Education, Culture, Sports, Science and Technology of Japan.  
\end{acknowledgments}

\vspace{0.5cm}
\appendix*

\section{A. Derivation of the physical temperature}
\renewcommand{\theequation}{A\arabic{equation}}
\setcounter{equation}{0}

\subsection{$q$-average}
We will derive the physical temperature, defining
\begin{eqnarray}
X_q &=& {\rm Tr} \:\left[1-(1-q) \gamma 
\left(H-U_q \right) \right]_+^{1/(1-q)}, \\
c_q &=& \frac{1}{(X_q)^q} \:{\rm Tr} \:\left[1-(1-q) \gamma 
\left(H-U_q \right) \right]_+^{q/(1-q)},
\label{eq:X1} 
\end{eqnarray}
where Tr denotes the trace.
We obtain \cite{Tsallis01}
\begin{eqnarray}
c_q &=& (X_q)^{1-q} = \nu_q (Z_{q})^{1-q}.
\end{eqnarray}
The physical temperature $T$ is assumed to be given by \cite{Abe01f}
\begin{eqnarray}
\frac{1}{T} &=& \frac{1}{c_q}\frac{\partial S_q}{\partial U_q},\nonumber \\
&=& \frac{k_B}{X_q} 
\:\left(\frac{\partial X_q}{\partial U_q} \right), \nonumber \\
&=& k_B \: \gamma,
\label{eq:X3} 
\end{eqnarray}
yielding
\begin{eqnarray}
\gamma &=& \frac{1}{k_B T} = \beta,
\end{eqnarray}
where we employ the relation: 
$\partial X_q /\partial U_q= \gamma \: X_q $.

\subsection{Normal average}
For the normal average, we define $\tilde{X}_q$ and 
$\tilde{c}_{q}$ by
\begin{eqnarray}
\tilde{X}_q &=& {\rm Tr} \:\left[1-(q-1) \tilde{\gamma} 
\left(H-\tilde{U}_q \right) \right]_+^{1/(q-1)}, \\
\tilde{c}_q &=& \frac{1}{(\tilde{X}_q)^q} 
\: {\rm Tr} \: \left[1-(q-1) \tilde{\gamma} 
\left(H -\tilde{U}_q \right) \right]_+^{q/(q-1)}.
\end{eqnarray}
A simple calculation leads to
\begin{eqnarray}
\tilde{c}_q (\tilde{X}_q)^{q-1} &=& 
\frac{1}{\tilde{X}_q} \:{\rm Tr} \: \left[1-(q-1) \tilde{\gamma} 
\left(H- \tilde{U}_q \right) \right]_+^{q/(q-1)}, \nonumber \\
&=& \frac{1}{\tilde{X}_q} 
\:{\rm Tr} \:\{ \left[1-(q-1) \tilde{\gamma} 
\left(H- \tilde{U}_q \right) \right]_+^{1/(q-1)}
\left[1-(q-1) \tilde{\gamma} 
\left(H- \tilde{U}_q \right) \right]_+ \}, \nonumber \\
&=& 1,
\label{eq:Y1}
\end{eqnarray}
where we employ the definition of $\tilde{U}_q$ given by
\begin{eqnarray}
\tilde{U}_q &=& \frac{1}{\tilde{X}_q} 
\:{\rm Tr} \:\{ \left[1-(q-1) \tilde{\gamma} H \right]_+^{1/(q-1)} H \}.
\end{eqnarray}
Equation (\ref{eq:Y1}) leads to
\begin{eqnarray}
\tilde{c}_q &=& (\tilde{X}_q)^{1-q}
= (\tilde{\nu}_q)^{-1} (\tilde{Z}_{q})^{1-q}.
\label{eq:Y2}
\end{eqnarray} 
With the use of Eqs. (\ref{eq:Y2}) and (\ref{eq:D11}), the physical temperature
is given by \cite{Abe01f}
\begin{eqnarray}
\frac{1}{T} &=& \frac{1}{\tilde{c}_q}
\frac{\partial S_q}{\partial \tilde{U}_q}, \nonumber \\ 
&=& \left( \frac{k_B}{\tilde{X}_q} \right) 
\:\frac{\partial \tilde{X}_q}{\partial \tilde{U}_q } , \nonumber \\
&=& q\: k_B \:\tilde{\gamma},
\label{eq:Y3}
\end{eqnarray}
leading to
\begin{eqnarray}
\tilde{\gamma} &=& \frac{\beta}{q},
\end{eqnarray}
where we employ the relation: 
$\partial \tilde{X}_q/\partial \tilde{U}_q=q \:\tilde{\gamma}\:\tilde{X}_q$.

\section{B. Evaluations of averages by the exact approach}
\renewcommand{\theequation}{B\arabic{equation}}
\setcounter{equation}{0}

\subsection{The $q$-average}
We first discuss evaluations by the $q$-average given by
\begin{eqnarray}
Z_q(\alpha) &= & {\rm Tr} \: \left[1-(1-q)  
\alpha \:H \right]^{1/(1-q)} , \\
Q_q(\alpha) &=& \frac{1}{\nu_q Z_q(\alpha)}
{\rm Tr} \:\{ \left[1-(1-q) \alpha \:H \right]^{q/(1-q)}\:Q \}, 
\end{eqnarray}
by using the exact expressions for the gamma function 
\cite{Prato95,Rajagopal98}:
\begin{eqnarray}
y^{-s} &=& \frac{1}{\Gamma(s)} \int_0^{\infty} u^{s-1}e^{-yu}\:du 
\hspace{2cm}\mbox{for $s > 0$}, 
\label{eq:Z3}\\
y^s &=&\frac{i}{2 \pi} \Gamma(s+1) \int_C (-t)^{-s-1} e^{-yt}\:dt
\hspace{1cm}\mbox{for $s > 0$},
\label{eq:Z4}
\end{eqnarray}
where $Q$ denotes an arbitrary operator and $C$ the Hankel path in the complex plane. 
We obtain \cite{Prato95,Rajagopal98}
\renewcommand{\arraystretch}{1.5}
\begin{eqnarray}
Z_q(\alpha)
&=& \left\{ \begin{array}{ll}
\frac{1}{\Gamma\left(\frac{1}{q-1} \right)} 
\int_0^{\infty} u^{1/(q-1)-1} e^{-u}
Z_1[(q-1) \alpha u] \: du
\quad & \mbox{for $q > 1.0$}, \\
\frac{i}{2 \pi}\Gamma\left(\frac{1}{1-q}+1 \right) 
\int_C (-t)^{-1/(1-q)-1} e^{-t}
Z_1[-(1-q) \alpha t] \: dt
\quad & \mbox{for $q <  1.0$}, 
\end{array} \right. \nonumber \\
&& \label{eq:Z1}
\end{eqnarray}
\begin{eqnarray}
Q_q(\alpha)
&=& \left\{ \begin{array}{ll}
\frac{1}{\nu_q\: Z_q(\alpha)
\Gamma\left(\frac{q}{q-1} \right)} \int_0^{\infty} u^{q/(q-1)-1} e^{-u}
Z_1[(q-1) \alpha u] \:Q_1[(q-1) \alpha u] \: du 
\hspace{0cm} & \mbox{for $q > 1.0$}, \\
\frac{i}{2 \pi \nu_q\: Z_q(\alpha)}
\Gamma\left(\frac{q}{1-q}+1 \right) \int_C (-t)^{- q/(1-q)-1} e^{-t}
Z_1[-(1-q) \alpha t] \:Q_1[-(1-q) \alpha t]\: dt, \\ 
\hspace{0cm} & \mbox{for $q <  1.0$}, 
\end{array} \right. \nonumber \\
&& \label{eq:Z2}
\end{eqnarray}
where
\begin{eqnarray}
Z_1(\alpha) &=& {\rm Tr}\: e^{-\alpha \:H}, 
\label{eq:Z5} \\
Q_1(\alpha) &=& \frac{1}{Z_1(\alpha)}
{\rm Tr} \:\{e^{-\alpha \;H}\:Q \}
\label{eq:Z6}
\end{eqnarray}
and $\nu_q$ is given by Eq. (\ref{eq:C7}) and (\ref{eq:E7b}) 
for ideal gases and harmonic oscillators, respectively.

\subsection{The normal average}
Next we discuss evaluations by the normal average given by
\begin{eqnarray}
\tilde{Z}_q(\tilde{\alpha}) &=&{\rm Tr} \: \left[1-(q-1)  
\tilde{\alpha} \:H \right]^{1/(q-1)}, \\
\tilde{Q}_q(\tilde{\alpha}) &=& \frac{1}{\tilde{Z}_q(\tilde{\alpha})} 
{\rm Tr} \:\{ \left[1-(q-1) \alpha \:H \right]^{1/(q-1)}\:Q\}.
\end{eqnarray}
By using Eqs. (\ref{eq:Z3}) and (\ref{eq:Z4}), we obtain
\begin{eqnarray}
\tilde{Z}_q(\tilde{\alpha})
&=& \left\{ \begin{array}{ll}
\frac{i}{2 \pi}\Gamma\left(\frac{1}{q-1}+1 \right) 
\int_C (-t)^{-1/(q-1)-1} e^{-t}
\:Z_1[-(q-1) \tilde{\alpha} t] \: dt
\quad & \mbox{for $q > 1.0$}, \\
\frac{1}{\Gamma\left[1/(1-q) \right]} 
\int_0^{\infty} u^{1/(1-q)-1} e^{-u}
\:Z_1[(1-q) \tilde{\alpha} u] \: du
\quad & \mbox{for $q <  1.0$}, 
\end{array} \right. \nonumber \label{eq:Z7}\\
&& 
\end{eqnarray}
\begin{eqnarray}
\tilde{Q}_q(\tilde{\alpha}) 
&=& \left\{ \begin{array}{ll}
\frac{i \:\Gamma[1/(q-1)+1]}{2 \pi \tilde{Z}_q(\tilde{\alpha})}
\int_C (-t)^{-1/(q-1)-1} e^{-t}
\:Z_1[-(q-1) \tilde{\alpha} t]\: Q_1[-(q-1) \tilde{\alpha} t]\: dt
\quad & \mbox{for $q > 1.0$}, \\
\frac{1}{\tilde{Z}_q(\tilde{\alpha}) \Gamma\left[1/(1-q) \right]} 
\int_0^{\infty} u^{1/(1-q)-1} e^{-u}
\:Z_1^{(N)}[(1-q) \tilde{\alpha} u]\: Q_1[(1-q)\tilde{\alpha} t] \: du
\quad & \mbox{for $q <  1.0$}, 
\end{array} \right. \nonumber \label{eq:Z8} \\
&& 
\end{eqnarray}
where $Z_1(\alpha)$ and $U_1(\alpha)$ are given by Eqs. (\ref{eq:Z5}) 
and (\ref{eq:Z6}), respectively.

\newpage


\newpage
\begin{figure}
\begin{center}
\end{center}
\caption{
(Color online)
$S_q/k_B N$ of ideal gases as a function of $q$ for $N=100$ calculated 
by the $q$ average ($q$-av.: the dashed curve) and
the normal average (N-av.: the solid curve), 
and the difference between the two entropies
(the bold curve) ($D=1$, $T/T_0=1.0$: $T_0=h^2/k_B m$).
The inset shows the validity range of the two averages in the $q$-$N$ space:
the MEMs with the $q$- and normal averages are valid
for $0 < q < q_U$ and $q > q_L$, respectively, and for $q_L < q < q_U$
both the averages are valid: the dashed line denotes $N=100$.
}
\label{fig1}
\end{figure}

\begin{figure}
\begin{center}
\end{center}
\caption{
(Color online) $S_q/k_B N$ as functions of $q$ and $N$ obtained 
by the $q$-average ($D=1$, $T/T_0=1.0$): 
note the logarithmic scale in the ordinate and abscissa for $N$.
}
\label{fig2}
\end{figure}

\begin{figure}
\begin{center}
\end{center}
\caption{
(Color online) $S_q/k_B N$ as functions of $q$ and $T$ obtained by the $q$-average
($D=1$, $N=100$): note the logarithmic scale in the ordinate.
}
\label{fig3}
\end{figure}

\begin{figure}
\begin{center}
\end{center}
\caption{
(Color online) $S_q/k_B N$ as functions of $T$ and $N$ 
obtained by the $q$-average ($D=1$, $q=0.8$): 
note logarithmic scales in the ordinate and abscissa for $N$.
}
\label{fig4}
\end{figure}

\begin{figure}
\begin{center}
\end{center}
\caption{
(Color online) $\tilde{S}_q/k_B N$ as functions of $q$- and $N$ 
obtained by the normal average
($D=1$, $T/T_0=1.0$): note the logarithmic scale in the abscissa for $N$.
}
\label{fig5}
\end{figure}

\begin{figure}
\begin{center}
\end{center}
\caption{
(Color online) $\tilde{S}_q/k_B N$ as functions of $q$ and $T$ 
obtained by the normal average ($D=1$, $N=100$). 
}
\label{fig6}
\end{figure}

\begin{figure}
\begin{center}
\end{center}
\caption{
(Color online) $\tilde{S}_q/k_B N$ as functions of $T$ and $N$
obtained by the normal average
($D=1$, $q=1.2$): note the logarithmic scale in the abscissa for $N$.
}
\label{fig7}
\end{figure}

\begin{figure}
\begin{center}
\end{center}
\caption{
(Color online) The relative fluctuation of $\Delta U_q/U_q$
as functions of $q$ and $(\eta DN)$
calculated by the $q$ average: $\eta=1/2$ ($\eta=1$) for ideal gases
(harmonic oscillators).
}
\label{fig8}
\end{figure}


\begin{thebibliography}{99}

\bibitem{Tsallis88}C. Tsallis:
J. Stat. Phys. {\bf 52}, 479 (1988).

\bibitem{Tsallis98}C. Tsallis, R. S. Mendes, 
and A. R. Plastino: Physica A {\bf 261}, 534 (1998). 

\bibitem{Tsallis01}C. Tsallis,
in {\it Nonextensive Statistical Mechanics and Its Application},
edited by S. Abe and Y. Okamoto (Springer-Verlag, Berlin, 2001), p 3.

\bibitem{Tsallis04}C. Tsallis:
Physica D {\bf 193}, 3 (2004).

\bibitem{Curado91}E. M. F. Curado and C. Tsallis,
J. Phys. A {\bf 24} (1991) L69; {\bf 24}, 3187 (1991) (corrigenda);
{\bf 25}, 1019 (1992).

\bibitem{Martinez00}S. Martinez, F. Nicolas, F. Pennini,
and A. Plastino,
Physica A {\bf 286}, 489 (2000).

\bibitem{Tsallis09}C. Tsallis and U. Tirnakli,
J. Phys. Conf. Ser. {\bf 201}, 012001 (2010).

\bibitem{Tsallis05}C. Tsallis, M. Gell-Mann, and Y. Sato, 
Proc. Natl. Acad. Sc. USA {\bf 102}, 15377 (2005).

\bibitem{Caruso07}F. Caruso and C. Tsallis, 
Phys. Rev. E {\bf 78}, 021102 (2008).

\bibitem{Wilk07}G. Wilk and Z.Wlodarczyk, 
Physica A {\bf 376}, 279 (2007).

\bibitem{Nonext}Lists of many applications of the nonextensive
statistics are available at 
URL: \\
(http://tsallis.cat.cbpf.br/biblio.htm)
\bibitem{Plastino94}A. R. Plastino, A. Plastino, and C. Tsallis, 
J. Phys. A: Math. Gen. {\bf 27} (1994) 5707.

\bibitem{Prato95}D. Prato, 
Phys. Lett. A {\bf 203} (1995) 165.

\bibitem{Abe99}S. Abe, 
Phys. Lett. A {\bf 263} (1999) 424; Erratum {\bf 267} (2000) 456.

\bibitem{Abe01}S. Abe, S. Martinez, F. Pennini, and A. Plastino,
Physics Letters A {\bf 278} (2001) 249.

\bibitem{Martinez00b}S. Martinez, F. Pennin, A. Plastino, 
Phys. Lett. A {\bf 278} (2000) 47.

\bibitem{Abe01b}S. Abe, S. Martinez, F. Pennini, A. Plastino, 
Phys. Lett. A {\bf 281} (2001) 126.

\bibitem{Liyan08}L. Liyan and D. Jiulin,
Physica A {\bf 387}, 5417 (2008).
\bibitem{Lenzi01}E. K. Lenzi, R. S. Mendes, L. R. da Silva, and L. C. Malacarne,
Physica A {\bf 289} (2001) 44.

\bibitem{Lenzi02}E. K. Lenzi, M. K. Lenzi, H. Belich and L. S. Lucenac
Phys. Lett. A {\bf 292}, 315 (2002). 


\bibitem{Abe02}S. Abe,
Phys. Rev. E {\bf 66}, 046134 (2002).

\bibitem{Abe05}S. Abe and G. B. Bagci,
Phys. Rev. E {\bf 71}, 016139 (2005). 
 
\bibitem{Abe06}S. Abe,
Astrophysics and Space Science {\bf 305}, 241 (2006).

\bibitem{Abe08}S. Abe,
Europhys. Lett. {\bf 84}, 60006 (2008).
%
\bibitem{Abe09}S. Abe,
Phys. Rev. E {\bf 79}, 041116 (2009).

\bibitem{Abe09b}S. Abe,
J. Stat. Mech. P07027 (2009) [arXiv:0906.2908].

\bibitem{Ven10}R. C. Venkatesan, A. Plastino,
Physica A {\bf 389}, 1159 (2010).

\bibitem{Hanel09}R. Hanel, S. Thurner, and C. Tsallis,
Europhys. Lett. {\bf 85}, 20005 (2009).

\bibitem{Lutsko09}J. F. Lutsko, J. P. Boon, and P. Grosfils,
Europhys. Lett. {\bf 86}, 40005 (2009).

\bibitem{Wilk00}G. Wilk and Z. Wlodarczyk,
Phys. Rev. Lett. {\bf 84}, 2770 (2000).

\bibitem{Beck01}C. Beck,
Phys. Rev. Lett. {\bf 87}, 180601 (2001).

\bibitem{Beck05}C. Beck and E. G. D. Cohen,
Physica A {\bf 322}, 267 (2003).

\bibitem{Beck07}C. Beck,
in {\it Anomalous Transport: Foundations and Applications},
edited by G. Radons, R. Klages and I. M. Sokolov (Wiley, New York, 2008).


\bibitem{Tsallis03}C. Tsallis and A. M. C. Souza,
Phys. Rev. E {\bf 67}, 026106 (2003).

\bibitem{Crooks07}G. E. Crooks,
Phys. Rev. E {\bf 75} (2007) 041119.

\bibitem{Abe07}S. Abe, C. Beck, and E. G. D. Cohen,
Phys. Rev. E {\bf 76}, 031102 (2007).

\bibitem{Straeten08}E. Van der Straeten and C. Beck,
Phys. Rev. E {\bf 78}, 051101 (2008).



\bibitem{Abe01f}S. Abe, S. Martinez, F. Pennini, and A. Plastino,
Phys. Lett. A {\bf 281}, 126 (2001).

\bibitem{Abe01c}S. Abe,
Physica A {\bf 300}, 417 (2001)

\bibitem{Abe01e}S. Abe, 
Phys. Rev. E {\bf 63}, 061105 (2001).

\bibitem{Abe01d}S. Abe, S. Martinez, F. Pennini, and A. Plastino,
Phys. Lett. A {\bf 281}, 126 (2001).

\bibitem{Hasegawa05}H. Hasegawa,
Physica A {\bf 351}, 273 (2005).

\bibitem{Abe06b}S. Abe,
Physica A {\bf 368}, 430 (2006).


\bibitem{Abe99b}S. Abe, 
Physica A {\bf 269}, 403 (1999).


\bibitem{Feng10}Zhi-Hui Feng and Li-Yan Liu,
Physica A {\bf 389}, 237 (2010).


\bibitem{Touch02}H. Touchette,
{\it Nonextensive Entropy-Interdisciplinary Applications},
ed. M. Gellmann and C. Tsallis (Oxford Univ. Press), p. 159.

\bibitem{Hasegawa10}H. Hasegawa,
J. Math. Phys. (in press) [arXiv:1001.0214].


\bibitem{Rajagopal98}A. K. Rajagopal, R. S. Mendes and E. K. Lenzi, 
Phys. Rev. Lett. {\bf 80}, 3907 (1998).

\end{thebibliography}
\end{document}